\documentclass[fleqn,10pt]{wlscirep}
\usepackage[utf8]{inputenc}
\usepackage[T1]{fontenc}
\usepackage{bm}
\usepackage{graphicx}
\usepackage{sidecap}
\usepackage{subcaption}
\usepackage{mwe}
\usepackage{wrapfig}
\usepackage{float}
\usepackage{caption}
\usepackage{xcolor}
\usepackage{soul}
\usepackage{amsmath}
\usepackage{gensymb}

\title{Heart rate measurement using the built-in triaxial accelerometer from a commercial digital writing device}

\author[1]{Julie Payette}
\author[1]{Fabrice Vaussenat}
\author[1,*]{Sylvain G. Cloutier}

\affil[1]{École de technologie supérieure, Department of Electrical Engineering, Montréal, H3C 1K3, Canada}

\affil[*]{SylvainG.Cloutier@etsmtl.ca}

\begin{abstract}
Wearable devices are on the rise. Smart watches \& phones, fitness trackers or smart textiles now provide unprecedented access to our own personal data. As such, wearable devices can enable health monitoring without disrupting our daily routines. In clinical settings, electrocardiograms (ECGs) and photoplethysmographies (PPGs) are used to monitor the heart's and respiratory behaviors. In more practical settings, accelerometers can be used to estimate the heartrate when they are attached to the chest. They can also help filter out some noise in ECG signal from movement. In this work, we compare the heart rate data extracted from the built-in accelerometer of a commercial smart pen equipped with sensors (STABILO's DigiPen), with a standard ECG monitor readouts. We demonstrate that it is possible to accurately predict the heart rate from the smart pencil. The data collection is done with eight volunteers, writing the alphabet continuously for five minutes. The signal is processed with a Butterworth filter to cut off noise. We achieve a mean-squared error (MSE) better than 6.685x$10^{-3}$ comparing the DigiPen's computed $\Delta t$ (time between pulses) with the reference ECG data. The peaks' timestamps for both signals all maintain a correlation higher than 0.99. All computed heart rates (HR =$\frac{60}{\Delta t}$) from the pen accurately correlate with the reference ECG signals. 
\end{abstract}

\begin{document}
\flushbottom
\maketitle

\thispagestyle{empty}

\section*{Introduction}

Medical devices have become part of our daily lives, especially in the wellness field\cite{tsoukas_review_2021,iqbal_advances_2021}. In the  Internet of things (IoT) era, systems of wireless, interconnected and networked digital devices can now continuously collect, analyse, send and store data over a network, making it tremendously easier to monitor a patient in real time in his living environment\cite{kelly_internet_2020}. These wearable devices use embedded accuracy sensors to non-invasively monitor healthcare data\cite{subahi_edge-based_2019} from the wrist\cite{haescher_study_2015}, the chest\cite{hung_estimating_2017}, the fingers\cite{alajlan_tinyml_2022,pham_wearable_2020} or other parts of the body. The main idea is to perform better medical monitoring during daily activities with precision and safety, all while preserving the patient's privacy and quality of life. Coupling these wellness devices with emerging data science and machine learning capabilities can provide a more accurate and more comprehensive dashboard of medical indicators\cite{cruz-ramos_mhealth_2022}. This can significantly reduce medical risks by including the patient in the healthcare management loop as early as possible\cite{lu_wearable_2020}. At the same time, wearable medical devices can simplify the monitoring of patients and keep them at home to improve their quality of life. The improved sensors,  micro-controllers, and end-point computing framework architectures unleash formidable possibilities for qualitative and predictive patient monitoring\cite{tsoukas_review_2021}. To facilitate the IoT integration in daily lives, a new approach also consists of measuring medical information using smart clothes\cite{bryson_20_2023}, watches\cite{yang_spo2_2015}, apps and common instruments like pens\cite{noauthor_digital_nodate}. The establishment of an easy-to-use surveillance system is an interesting path, especially considering the aging population, the prevalence of cardiovascular disease\cite{wimmer_clinical_2013},the emergence of infectious diseases\cite{clerkin_covid-19_2020} affecting the cardiovascular system and psychological diseases leading to behavioral disorders\cite{schneiderman_stress_2005}. Therefore, different medical devices that can measure physiological parameters such as ECG, heartbeat derived from ECG, respiratory rate (RR), SPO2\cite{tamura_current_2019}, temperature\cite{boano_accurate_2011,basra_temperature_2017}, blood pressure\cite{chang_cuff-less_2019} are vital to monitor fragile patients daily. The choice of sensors is based on the bio-signals to be monitored. Capacitance electrodes are commonly used to detect ECG\cite{lim_capacitive_2014}, respectively inductance electrodes are used to measure breathing information\cite{shen_respiratory_2017}. Studies show that sets of accelerometers can follow the ribcage movements to accurately predict the respiratory rate as well as the ECG\cite{chan_ambulatory_2013}. Some of these sensors can monitor multiple medical parameters simultaneously\cite{pandian_smart_2008}.  Most of these wearable multi-parameter medical devices are used to detect heart and lung disorders\cite{iqbal_advances_2021}. \\

\noindent The literature already shows that respiratory rate and heartbeat can be inferred from ECG signals\cite{sarkar_extraction_2015,mirmohamadsadeghi_respiratory_2014}. Similarly, blood pressure can be measured using PPGs\cite{wu_flexible_2022}  and ECGs\cite{chang_cuff-less_2019} allowing for a cuffless blood pressure monitoring that is more comfortable for the patient and can be integrated into a biosensor frame. PPG is a non-invasive optical technique from which many physiological parameters can be derived, as it produces a waveform that correlates with circulatory volume in skin tissue. SPO2 is mainly used to measure PPG using infrared optical sensors usually located on the forehead, finger and more recently on the rib cage \cite{moco_new_2018}. In addition, one of the main applications is to detect dysautonomia and to monitor central nervous system imbalance to assess patients' behavioral disorders like stress \cite{owens_psychophysiology_2016}. Dysautonomia\cite{leti_interets_nodate} describes the  imbalance between the sympathetic and parasympathetic nervous system that may be present in anxiety-inducing situations and more generally in behavioral disorders.\cite{tulba_dysautonomia_2020,lo_covid-19_2021,goldstein_dysautonomia_2014}. It is based on heart rate variability (HRV), which is calculated from the heart rate using the respiratory rate (RR) intervals\cite{shahani_rr_1990}. This imbalance caused by noxious stimuli can stress homeostasis\cite{mckendrick_familial_1958}. Chronic stress causes physical, psychological, and behavioral abnormalities by hyperactivating the sympathetic nervous system and deteriorate the patient's condition. It is accepted that HRV is a good method to monitor and assess the stress status\cite{jovanov_stress_2003}. HRV represents the heart's ability to respond to a variety of physiological and environmental stimuli\cite{rajendra_acharya_heart_2006}. Low HRV is associated with a monotonous, regular heart rate. Furthermore, low HRV has also been associated with impaired regulatory and homeostatic functions of the ANS, which reduces the body's ability to cope with internal and external stressors. In fact, we can see that an accelerometers are often used especially to detect ECG, heartbeat and respiratory rate especially located on the wrist to compensate for motion artifacts or the ribcage using respiratory movements\cite{zhao_robust_2021}.\\

\noindent In this study, we demonstrated that the heart rate can be accurately monitored using the built-in accelerometer within a digital smart-pen (DigiPen©) during the writing process\cite{noauthor_digitizing_nodate,oberdorf_kit_2022}. For example, using a pen as a heart rate recording system may present a significant opportunity for monitoring students stress in the classroom . This could thereby provide a method of assessing the impact of stress on cognitive abilities\cite{yaribeygi_impact_2017}, and even potentially detect hyperactivity, attention issues or other cognitive and behavioral disorders\cite{hollocks_differences_2014,denckla_theory_1996}. Dysautonomia also plays a role in detecting cognitive disorders such as Alzheimer's and Parkinson's disease \cite{goldstein_dysautonomia_2014,rolinski_rem_2014}. As such, a simple act of writing could offer a means to evaluate these cognitive disorders.

\section*{Materials and Methods}

\subsection*{Data Collection}

\subsubsection*{Materials}

\begin{figure}[H]
     \centering
     \begin{subfigure}[b]{0.45\textwidth}
         \centering
         \includegraphics[width=\textwidth]{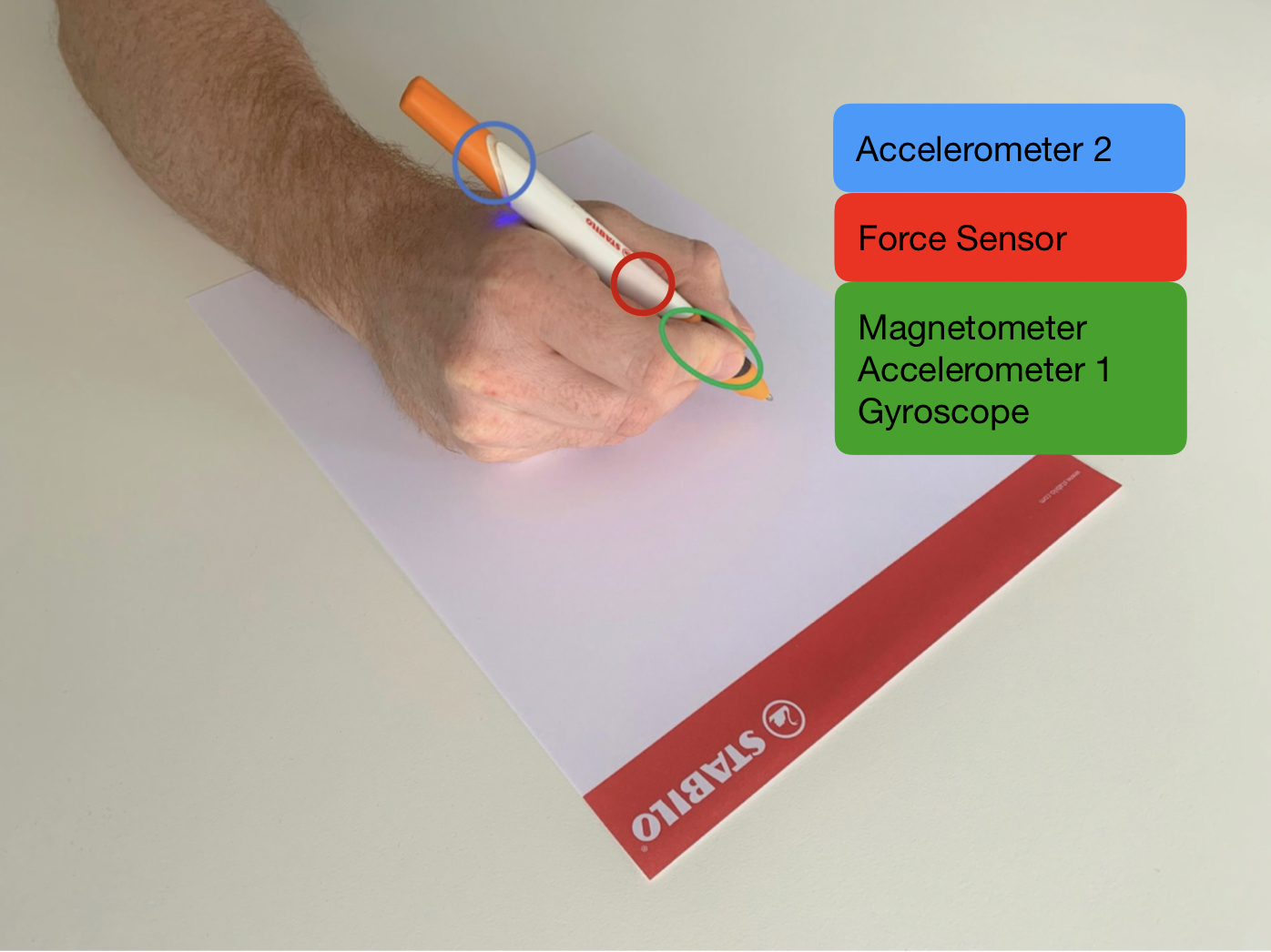}
         \caption{}

     \end{subfigure}
     \hfill
     \begin{subfigure}[b]{0.45\textwidth}
         \centering
         \includegraphics[width=\textwidth]{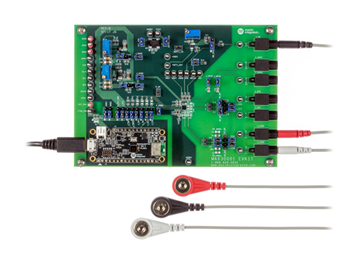}
         \caption{}

     \end{subfigure}
        \caption{(a) STABILO's Digipen sensors' location\cite{noauthor_sensors_nodate} (b) MAX3001 EV kit and electrodes\cite{integrated_max30001_nodate}}
        \label{fig1}
\end{figure}

\noindent For this study, we worked with the commercially-available DigiPen from STABILO\cite{noauthor_digipen_nodate}. It is a smart pen equipped with sensors, which is marketed as a tool for assessment of writing motor skills. The pen shown in Figure 1(a) features two built-in triaxial accelerometers (front and rear), a gyroscope, a magnetometer as well as a force sensor. All data is streamed directly to a connected device with the \textit{DigiPen : Development Kit Demo} app installed using BLE connection. In order to retrieve the heart rate, we use the accelerometer located near the tip of the pen, where the fingertips rest. It is a LSM6DSL accelerometer with a sampling rate of 100Hz\cite{noauthor_sensors_nodate}.
\\

\noindent To  evaluate the heart rate predictions from the accelerometer data, we use the MAX30001\cite{integrated_max30001_nodate} Evaluation System. It is an assembled circuit that also provides a platform to monitor the data and save it directly in your computer to gather the ECG signal. In our case, we more specifically recorded the heart pulses, which correspond to the RR intervals to calculate the cardiac rhythms. There are several components to an ECG signal. They can be separated into segments known as waves\cite{ashley_conquering_2004}.
\\

 \begin{figure*}[h]
        \centering      
        \includegraphics[width=0.75\textwidth]{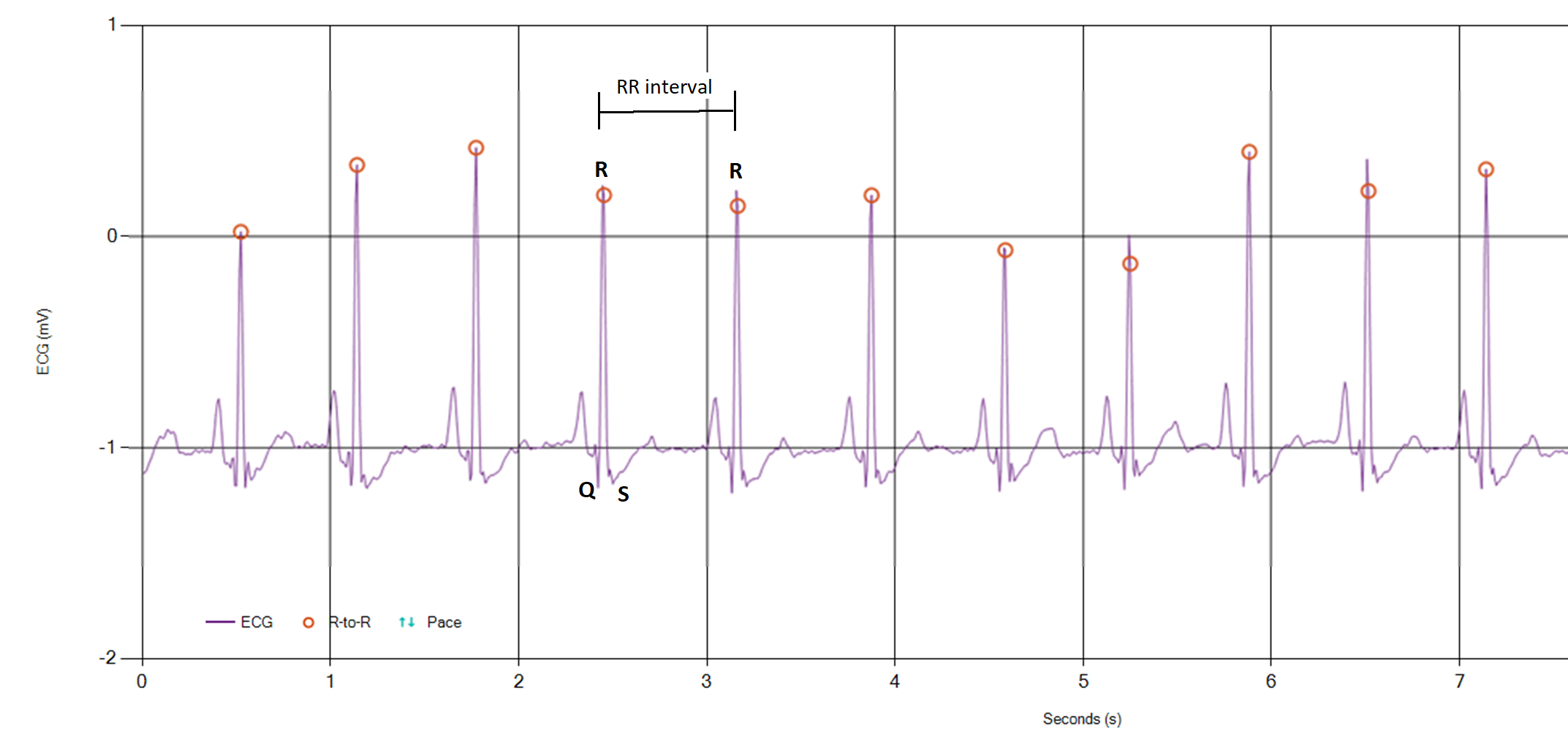}      
        \caption{A typical reference ECG signal recorded with the MAX30001 Evaluation system. Red dots are the detected R-waves. RR interval is shown as the time period between two R-waves. The QRS complex is the ventricular depolarization. }
        \label{fig2}
              
    \end{figure*}

\noindent The R-wave correspond to the heart beat. Figure 2 clearly shows the first positive deflection of the QRS complex, which is the ventricular depolarization\cite{wei_physiology_2023}. Indeed, the R-wave is the highest peak because it is the the ventricules' main mass that depolarizes\cite{ashley_conquering_2004}. Hence, by measuring the time interval between two consecutive R-waves (i.e. the RR interval) one can directly compute the heart rate as $\text{HR}=\frac{60}{{\Delta t}_{RR}}$\cite{prakash_how_2005}.\\

\noindent However, that RR interval can vary between peaks. As such, computing the HR from sequential intervals does not always give the same results. This is because the heart rate can change continuously throughout the day. Those fluctuations between heartbeats are the heart rate variability (HRV).\cite{zhu_heart_2019}
\\

\subsubsection*{Methods}
    \begin{figure*}[h]
        \centering      
        \includegraphics[width=0.55\textwidth]{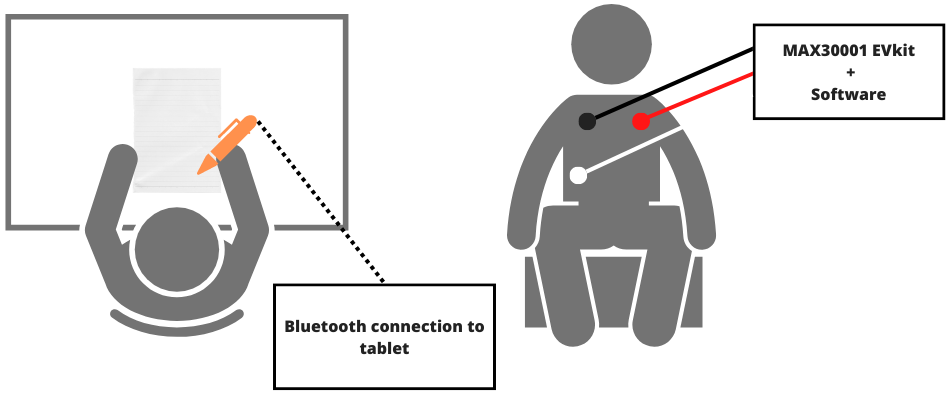}      
        \caption{Schematic Representation of our experimental configuration, top and front views. }
        \label{fig3}
              
    \end{figure*}

\noindent For the data acquisition, we selected eight (8) participants. All participants are in healthy condition. They wore the MAX30001 EV systems' electrodes on their chest to collect their ECG signal over a five-minute period. During this time, they were seated in a relaxed position and had to write the alphabet repeatedly on a sheet of paper using the DigiPen. They were asked to hold the pen in the right orientation, with the Stabilo logo facing upwards as shown in Figure 3. Four (4) male participants and four (4) female participants contributed to this preliminary data collection. Only one participant is left-handed. Since the left-handed population is estimated to represent 10-15\% of the world population\cite{sommer_language_2009}, this is deemed a representative sample.
\\

\noindent To process the signal from the accelerometer, we use Python on Google Colab. First, the anonymous raw data is re-scaled according to the sensor's specifications provided on STABILO's website\cite{noauthor_sensors_nodate}. Then, we combine the 3 separate axis to obtain the Eucledian acceleration trace, using the magnitude given as $a=\sqrt{{a_x}^2+{a_y}^2+{a_z}^2}$, where ${a_x}, {a_y}, {a_z}$ correspond to the accelerometer's three components.\\

\noindent In order to reduce the noise, we apply a fourth-order Butterworth\cite{butterworth_theory_1930} lowpass 2Hz filter to smooth out the signal, as shown in Figure 4. Finally, the goal is to align the detected peaks from the accelerometer's signal with the R-waves of the ECG signal. Four vectors are generated from the data. For both the ECG and DigiPen datasets, we create two vectors : (1) a time vector containing the timestamps and the moments of each peaks recorded over a five-minute interval and (2) a $\Delta$T vector containing the time differences between consecutive peaks. The dimension of these vectors obviously varies with the participants' heart rate. For a lower heart rate, fewer peaks are detected during a five-minute period than for a higher heartbeat. As seen previously, the heart rate is inversely proportional to the RR interval. As such, a larger RR interval means a lower heart rate.
\\
        \begin{figure*}[h]
        \centering      
        \includegraphics[width=0.6\textwidth]{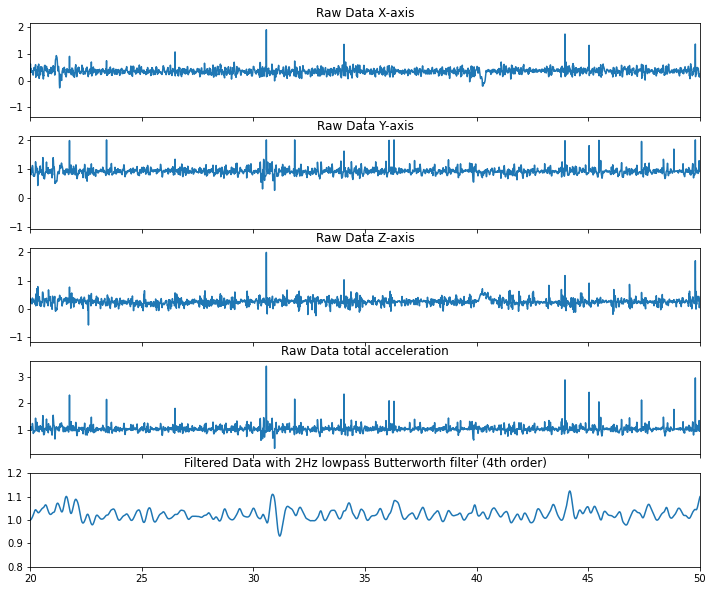}      
        \caption{Signal processing of the accelerometer data, plotted on a 30-second period for a more detailed view}
        \label{fig4}
              
    \end{figure*}

\section*{Results and Discussion}

As seen in Figure 5(a), the peaks' alignment is quite precise. While there is sometimes a delay of less than one second, it does not have a negative impact on our results since we use the time difference between two pulses to determine the heart rate.
\\

\begin{figure}[h!]
     \centering
     \begin{subfigure}[b]{0.48\textwidth}
         \centering
         \includegraphics[width=\textwidth]{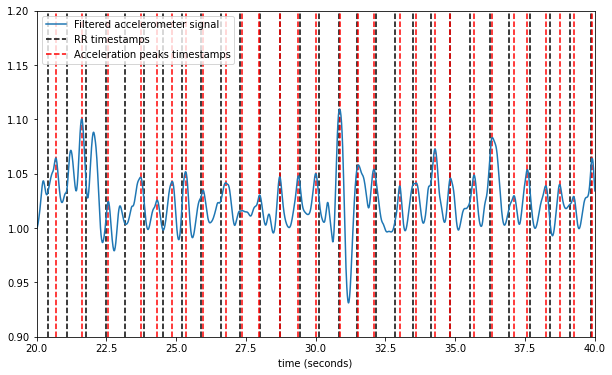}
         \caption{}

     \end{subfigure}
     \hfill
     \begin{subfigure}[b]{0.48\textwidth}
         \centering
         \includegraphics[width=\textwidth]{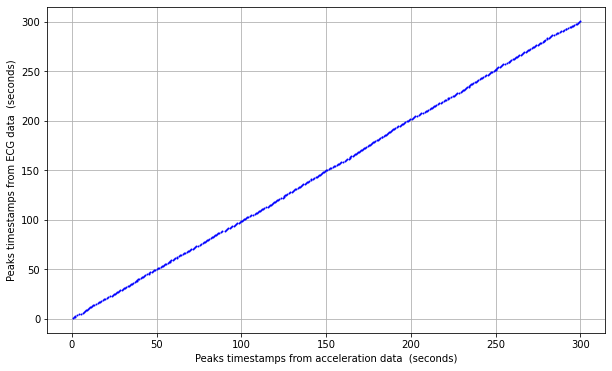}
         \caption{}

     \end{subfigure}
        \caption{(a) DigiPen's accelerometer signal after filtering with dotted vertical lines indicating the peaks' timestamps from the accelerometer (\textcolor{red}{-}) and the RR interval (-) from the ECG. It is plotted on a 20-second period for a more detailed view. (b) Correlation between the peaks' timestamps.}
        \label{fig5}
\end{figure}

\noindent Figures 5(a,b) shows the typical data obtained for one of our eight (8) participants, but they are highly representative of all our individual test results. Table 1 shows the regularity and consistency of our protocol. Despite the participants having different resting heart rates, we still manage to predict them accurately from the DigiPen's accelerometer data. 
\\

\begin{table}[h!]
  \caption{Results obtained from our eight (8) participants. Mean $\Delta$t and its standard deviation extracted from both the ECG and Digipen, as well as the mean heart rate. Pearson R and Cosine similarity factors are computed for the time vectors. In contrast, the Welch test's p-value is calculated from the $\Delta$t vectors.}
  \label{tbl1}
  \includegraphics[width=\linewidth]{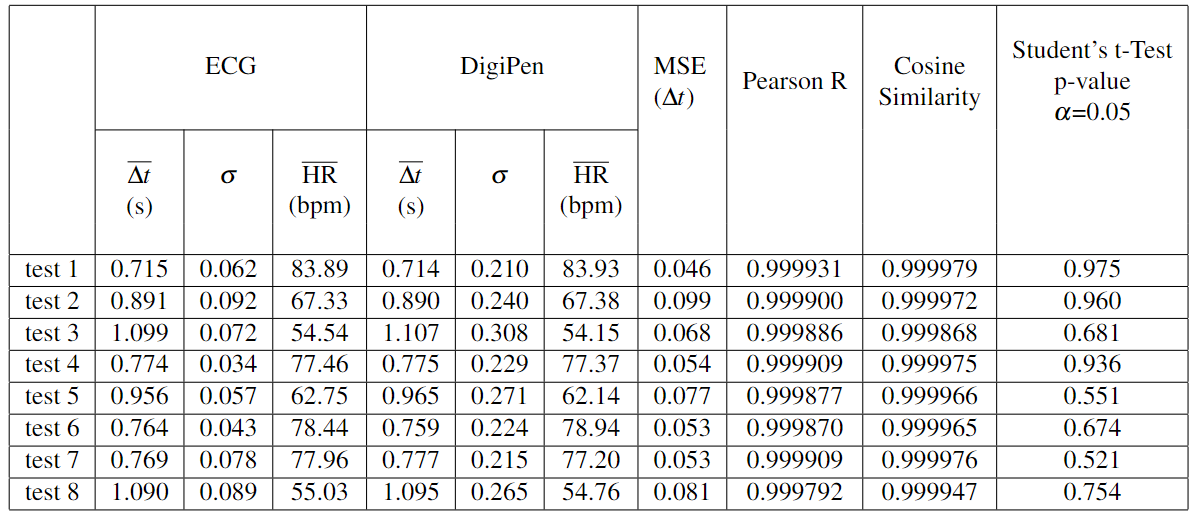}
\end{table}

\noindent As seen in Table 1, the time vectors are practically the same. Both the Pearson correlation coefficients and the cosine similarity factors are computed to suggest over 99\% correlation. Meanwhile, the derived average time difference $\overline{\Delta t}$ from the DigiPen data accurately correlates with the reference ECG data. Obviously, since the pen is not as precise as the ECG, the standard deviation for the DigiPen's results is always larger. However, this doesn't affect the similarity between the heart rates values. The most considerable deviation recorded is only 0.76 bpm, which is negligible for clinical purpose since the heart rate is usually expressed to the nearest unit.
\\

\noindent The filtering was optimized through comparison of peaks with the ECG signal. The data collected on the EVkit, which is the RR-interval, gives us the number of heartbeats during the test. To find a corresponding number from the accelerometer signal, we programmed a verification loop on Python that tests different frequency filters and computes the number of peaks to find the best match. The 2Hz frequency gave the best results in our case. Figure 7 shows a comparison for the peaks's correlation for three frequencies : half our optimal frequency (1Hz), the optimal frequency (2Hz) and its double (4Hz). Although a 1Hz-frequency filter results in a similar correlation, 2Hz gives a more precise one. For the 4Hz low-pass filter, we see that the correlation deteriorates rapidly. However, 2Hz is the lowest frequency filter we can use, otherwise the data loses too much detail and identifying the peaks becomes impossible.\\

    \begin{figure*}[h!]
        \centering      
        \includegraphics[width=0.6\textwidth]{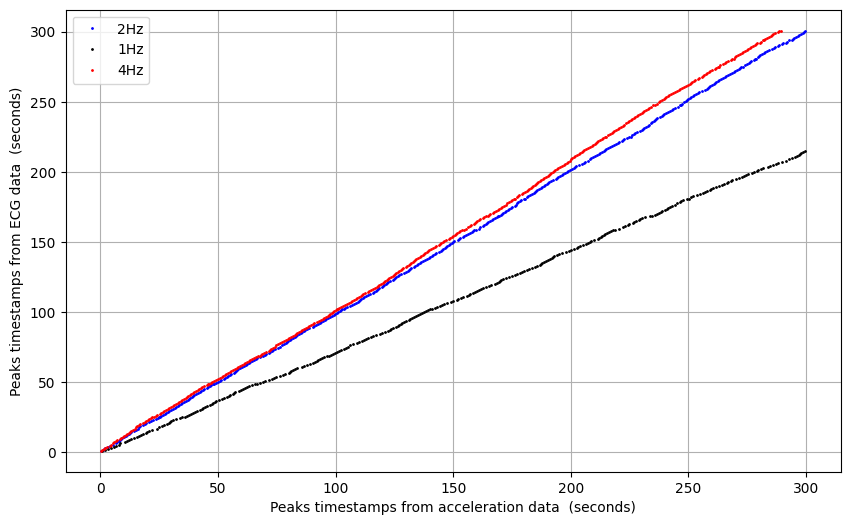}      
        \caption{Correlation comparison between the peaks' timestamps for frequency filters 1Hz(\textcolor{red}{\textbullet}), 2Hz(\textcolor{blue}{\textbullet}), 4Hz(\textcolor{black}{\textbullet})}
        \label{fig6}
              
    \end{figure*}
\noindent We also calculated the mean squared error  $(MSE =\frac{1}{n}  \sum_{j=1}^n (Y_{j}-\hat{Y}_{j})^2 )$ for the eight test's $\Delta t$ vectors. All are in a $10^{-2}$ range. Furthermore, we also performed Student's t-Test on these $\Delta t$-vectors. Every p-value is greater than 0.05, so the null hypothesis that both means are equal can be accepted for each of our eight tests. Finally, Figure 6 shows the comparison boxplots. Although wider spreads can be noticed using the pen's accelerometer data, the median values are all similar, with a maximal difference of 0.08. The skewness from the pen's data can be explained by some remaining noise in the signal.\\

    \begin{figure*}[h!]
        \centering      
        \includegraphics[width=\textwidth]{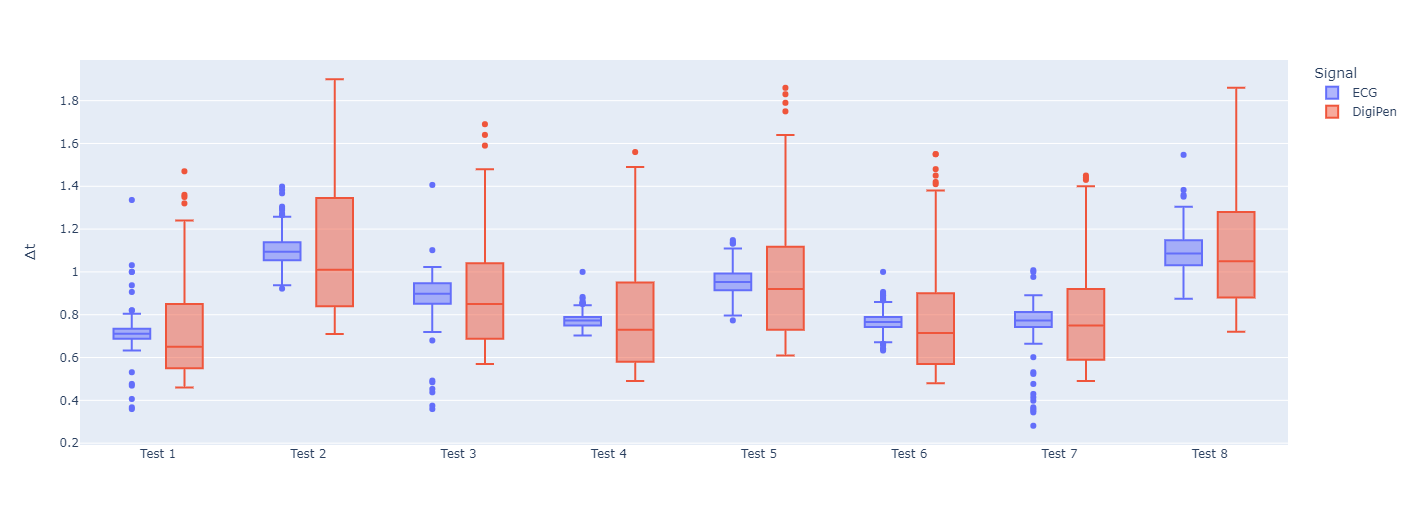}      
        \caption{Boxplots comparison for $\Delta$t-vectors from ECG vs DigiPen on all tests}
        \label{fig7}
              
    \end{figure*}

\section*{Conclusion}

In clinical settings, the patients' heart behaviors can and will continue to be monitored with cutting-edge equipment. On the other hand, this work shows the capability a smart pen's accelerometer to continuously and precisely determine the heart rate. Indeed, we managed to compute every participant's heartrate from the pen with a most impressive deviation of only 0.76bpm. This could prove useful for noninvasive and unbiased behavioural assessments, including heart issues like arrythmia\cite{berntson_heart_1997}, stress\cite{zhu_heart_2019} or fatigue\cite{da_estrela_heart_2020}. Since cardiac rhythms are modulated by the sympathetic and parasympathetic branches of the autonomic nervous system, HRV could also be useful for the early detection and monitoring of Parkinson's disease evolution in patients\cite{li_association_nodate,dorantes-mendez_characterization_2022}. Rightly so, future work includes a more detailed dataset and the development of a machine learning algorithm capable of identifying if the participant was under stress or not during the writing process.

\newpage
\bibliography{references}

\end{document}